\documentclass[prb,twocolumn,showpacs,preprintnumbers,amsmath,amssymb]{revtex4}
\usepackage{graphicx}
\usepackage{dcolumn}
\usepackage{bm}
\usepackage{natbib}
\usepackage[T1]{fontenc}

\begin{document}

\title{Physical meaning and measurement of the entropic parameter $q$ in an inhomogeneous magnetic systems}

\author{M.S. Reis}
\email{marior@fis.ua.pt}
\author{V.S. Amaral}
\affiliation{Departamento de F\'{i}sica and
CICECO, Universidade de Aveiro, 3810-193 Aveiro,
Portugal}
\author{R.S. Sarthour and I.S. Oliveira}
\affiliation{ Centro Brasileiro de Pesquisas
Físicas, Rua Dr. Xavier Sigaud 150 Urca,
22290-180 Rio de Janeiro-RJ, Brasil}

\date{\today}

\begin{abstract}

In this paper we present a thorough analysis of
two systems magnetically inhomogeneous: the
manganite La$_{0.7}$Sr$_{0.3}$MnO$_3$/MgO and the
amorphous alloy Cu$_{90}$Co$_{10}$. In both
cases, the non-extensive statistics yield a
faithful description of the magnetic behavior of
the systems. In the model proposed here, the
inhomogeneous magnetic system is composed by many
Maxwell-Boltzmann homogeneous bits and the
entropic parameter $q$ is related to the moments
of the distribution of the inhomogeneous
quantity. From the analysis of Scanning
Tunnelling Spectroscopy (STS) images, the $q$
parameter can be directly measured.

\end{abstract}

\maketitle

\section{Introduction}
 Tsallis thermostatistics has been widely used in a number
of different contexts\cite{website}. The framework is applicable
to systems which, broadly speaking, present at least one of the
following properties: (i) long-range interactions, (ii) long-time
memory, (iii) fractality and (iv) intrinsic inhomogeneity
\cite{website,livro_tsallis}. Manganese oxides, or simply
manganites, seems to embody three out of these four ingredients:
they present Coulomb long-range interactions
\cite{PRB_64_2001_235127,PRB_64_2001_235128,science_283_1999_2034},
clusters with fractal shapes
\cite{PR_344_2001_1,PRB_66_2002_174436} and intrinsic
inhomogeneity
\cite{PR_344_2001_1,livro_dagotto,PRL_89_2002_237203,PRL_92_2004_126602,nature_428_2004_401}.
Indeed, in a sequence of previous publications
\cite{EL_58_2002_42,PRB_66_2002_134417,prb_68_2003_014404}, it has
been shown that the magnetic properties of manganites can be
properly described within a mean-field approximation using Tsallis
statistics.

\section{PHYSICAL PICTURE}

We consider an inhomogeneous magnetic system composed by many
homogeneous parts (clusters) with different sizes, each one of
them described by the Maxwell-Boltzmann statistics. By averaging
the magnetization over the whole system, we recover the Tsallis
non-extensivity. A relationship between the $q$ parameter and the
moments of the distribution is obtained. The model is tested using
Scanning Tunnelling Spectroscopy (STS) conductance maps, where the
$q$ parameter could be obtained and, consequently, the bulk
magnetization predicted. We also apply the results to describe the
magnetic behavior of Cu$_{90}$Co$_{10}$ amorphous ribbons.

The starting point is a magnetic system formed by small regions,
or clusters of Maxwell-Boltzmann bits, each one of them with
magnetization $\mathcal{M}(\mu,T,H)$ given by the usual Langevin
function\cite{livro_APG}:
\begin{equation}\label{langevin}
\mathcal{M}(\mu,T,H)=\mu\left[\coth x-\frac{1}{x}\right]
\end{equation}
where $x=\mu H/kT$. The clusters are distributed in size, and
therefore in their net magnetic moment. Let $f(\mu)$ be the
distribution of the clusters magnetic moment. Thus, the average
magnetization of the sample will be given by:
\begin{equation}\label{mag_media_momento}
\langle\mathcal{M}\rangle=\int_0^\infty \mathcal{M}(\mu,T,H)f(\mu)
 d\mu
\end{equation}

On the other hand, Ref. \cite{prb_68_2003_014404,comentario1}
shows that the non-extensive magnetization is given by the
generalized Langevin function:
\begin{equation}\label{mag_generalizada}
\mathcal{M}_q=\frac{\mu_{ne}}{(2-q)}\left[\coth_qx-\frac{1}{x}\right]
\end{equation}
where $q\in\Re$ is the Tsallis entropic parameter and
$x=\mu_{ne}H/kT$.

Equating the susceptibilities
\begin{equation}\label{equating}
\chi_q=\langle\chi\rangle\equiv\lim_{H\rightarrow 0}\frac{\partial
M}{\partial H}
\end{equation}
and the saturation values ($H\rightarrow \infty$) of Eqs.
\ref{mag_media_momento} and \ref{mag_generalizada}, we find a
analytical expression to the $q$ parameter:
\begin{equation}\label{q_momento}
q(2-q)^2=\frac{\langle\mu^2\rangle}{\langle\mu\rangle^2}
\end{equation}
where $\langle\mu\rangle$ and $\langle\mu^2\rangle$ are the first
and second moments of the distribution $f(\mu)$, respectively.
This result is valid for any $f(\mu)$, and is analogous to that
obtained by Beck \cite{PRL_87_2001_180601} and Beck and Cohen
\cite{PA_322_2003_267}.

Now, let us consider a distribution of critical
temperatures. Again, the magnetization of a small
region is given by the usual Langevin function
(Eq.\ref{langevin}), however, within the
mean-field approximation, where:
\begin{equation}\label{x_mean_field}
x=\frac{\mu (H_0+\lambda \mathcal{M})}{kT}
\end{equation}
and $\lambda=3kT_C/\mu^2$ corresponds to the mean-field parameter.
Unfortunately, considering a distribution of critical
temperatures, the average magnetization
\begin{equation}\label{mag_media_tc}
\langle\mathcal{M}\rangle=\int_0^\infty
\mathcal{M}(\mu,T,H_0,T_C,\mathcal{M})f(T_C)dT_C
\end{equation}
can not be reduced in a similar fashion to what was done in the
case of a distribution of magnetic moments. Thus, a closed and
simple expression connecting the $q$ parameter and the moments of
the distribution of critical temperatures could not be obtained.
However, as will be discussed, the connection with experimental
results will provide some suggestions.

\section{Connections with experimental results}

\subsection{Distribution of magnetic moments}

As discussed before, the $q$ parameter is related to the moments
of the distribution of magnetic moments (Eq. \ref{q_momento}).
Thus, to measure the entropic parameter we need to measure,
\emph{a priori}, the distribution of magnetic moments. Below, we
will discuss how to extract $f(\mu)$ from some experimental data.

\subsubsection{Manganites}

Colossal magnetoresistance (CMR) effect\cite{livro_dagotto},
usually observed on manganites, has been proposed in terms of
intrinsic
inhomogeneities\cite{livro_dagotto,PR_344_2001_1,PRL_89_2002_237203},
which can lead to a formation of insulating and conducting domains
within a single sample, i.e., electronic phase separation in a
chemical homogeneous sample. The inhomogeneities alter the local
electronic and magnetic properties of the sample and should
therefore be visible via STS
\cite{nature_416_2002_518,science_285_1999_1540,PRL_89_2002_237203}
or Magnetic Force Microscopy (MFM)
\cite{science_276_1997_2006,science_298_2002_805}.

\begin{figure}
\begin{center}
\includegraphics[width=7cm]{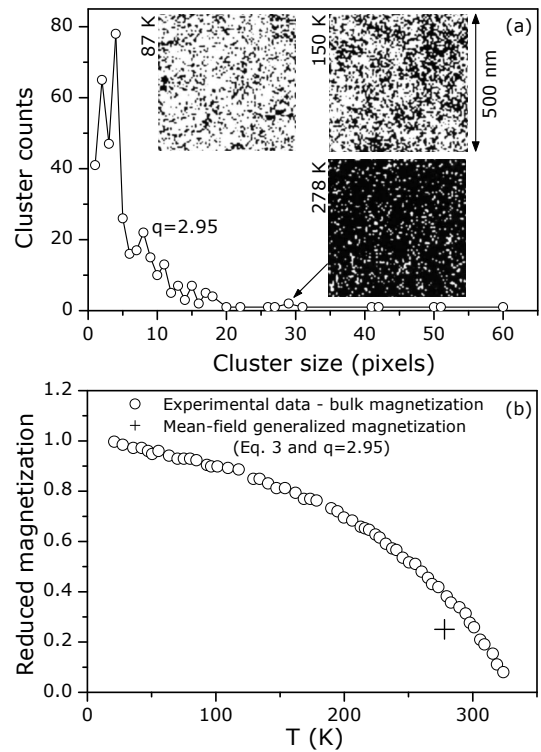}
\end{center}
\caption{(a) Scanning tunnelling spectroscopy
images obtained at different temperatures on
La$_{0.7}$Sr$_{0.3}$MnO$_3$/MgO manganite thin
film, after Becker \emph{et
al.}\cite{PRL_89_2002_237203}. White regions are
conducting (ferromagnetic phase) clusters and the
black regions are insulating clusters
(paramagnetic phase). The main graphic is the
cluster size distribution of the image at 278 K,
proportional to the cluster magnetic moment
distribution, where, using
Eq.\ref{relacao.mi.fi}, $q=2.95$ could be
directly obtained. (b) Using the mean-field
(reduced) generalized magnetization
(Eq.\ref{mag_generalizada}: $+$ symbol), we could
predict the bulk magnetization, in a satisfactory
agreement with the measured one ($\circ$ symbol),
in a reduced scale $\mathcal{M}/\mathcal{M}$(18
K)\cite{PRL_89_2002_237203}.} \label{reisfig1}
\end{figure}

Becker and co-workers\cite{PRL_89_2002_237203}
measured STS in a La$_{0.7}$Sr$_{0.3}$MnO$_3$/MgO
thin film and visualized a domain structure of
conducting (ferromagnetic) and insulating
(paramagnetic) regions with nanometric size,
since this manganite has a transition from a
metallic phase (below T$_C$) to an insulating
phase (above T$_C$), with a strong phase
coexistence/competition around T$_C\sim$ 330 K.
These STS conductance maps obtained by those
authors at 87 K, 150 K and 278 K are reproduced
in figure \ref{reisfig1}(a). From these 1-bit
images (black regions mean
insulating/paramagentic phase and white regions
stand for conducting/ferromagnetic phase), it was
possible to determine the distribution of
clusters size. Considering that the cluster size
$\phi$, measured in \emph{pixels}, is
proportional to the magnetic moment $\mu$ of the
cluster, Eq. \ref{q_momento} can be re-written
as:
\begin{equation}\label{relacao.mi.fi}
    \frac{\langle\phi^2\rangle}{\langle\phi\rangle^2}=\frac{\langle\mu^2\rangle}{\langle\mu\rangle^2}=q(2-q)^2
\end{equation}
The conductance map at 278 K has a distribution
of clusters as presented in figure
\ref{reisfig1}(a), and, using Eq.
\ref{relacao.mi.fi} we obtained from the data
$q=$2.95.

With this value of $q$, the total magnetization
of the system can be predicted, by considering
the mean-field approximation into the generalized
magnetization (Eq.\ref{mag_generalizada}), where
$x=3m_q/t$, $m_q=\mathcal{M}_q/\mu_{ne}$,
$t=T/T_C^{(1)}$ and $T_C^{(1)}=298
K$\cite{EL_58_2002_42} (see refs.
\cite{prb_68_2003_014404,EL_58_2002_42} for
details concerning the mean-field approximation
applied to the non-extensive magnetization). This
procedure results in a satisfactory agreement
between the predicted reduced magnetization
($m_q=$0.25; the $+$ symbol) and the experimental
one (the $\circ$ symbol), obtained measuring the
bulk magnetization\cite{PRL_89_2002_237203}, as
presented in figure \ref{reisfig1}(b). The images
at 87 K and 150 K were not analyzed, since the
clusters have already percolated.

The procedure above described shows how to extract the $q$
parameter from an experimental data, and then how to apply the
obtained $q$ parameter to predict macroscopic quantities of the
system. In addition, these results exemplify the relation between
non-extensivity and microscopic inhomogeneities. Finally, it is
important to stress that $q$ is related to the dynamics of the
system, since it measures the distribution of magnetic moments,
that contains the dynamics.

\subsubsection{Granular Alloys}

Ferrari and co-workers\cite{prb_56_1997_6086}
analyzed some melt-spun Cu$_{90}$Co$_{10}$
ribbons, materials intrinsically
inhomogeneous\cite{jnn_4_2004_1056,PRB_63_2001_014408}.
They considered Eq.\ref{mag_media_momento} to
study the magnetic behavior of these materials,
supposing a Log-Normal distribution of magnetic
moments:
\begin{equation}\label{lognormal}
    f(\mu)=\frac{1}{\sqrt{2\pi}\;s\;\mu}\;\exp\left[-\frac{\ln^2(\mu/\mu_0)}{2\;s^2}\right]
\end{equation}
where the k$^{th}$ moment is $\langle
\mu^k\rangle=\mu_0^k\exp(k^2s^2/2)$.

\begin{figure}
\begin{center}
\includegraphics[width=7cm]{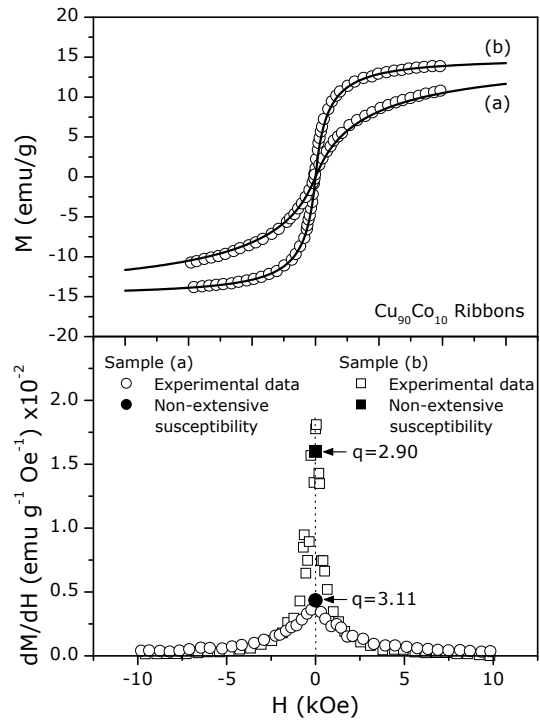}
\end{center}
\caption{Top: Open circles are the experimental
results, whereas the solid lines correspond to
the average magnetization
(Eq.\ref{mag_media_momento}), considering a
Log-Normal distribution (Eq.\ref{lognormal}).
Bottom: dM/dH as a function of the applied
magnetic field, where the magnetic susceptibility
lies in the zero field limit. The predicted
non-extensive magnetic susceptibility matches the
experimental one, for both samples.}
\label{reisfig2}
\end{figure}

Those authors used this model to fit the
magnetization curves at room temperature, as
displayed in figure \ref{reisfig2}-top. The
fitting parameters for the sample (a) are
$\mu_0=500\mu_B$ and $s=1.16$; and for the sample
(b), $\mu_0=3900\mu_B$ and $s=0.93$. Both samples
are Cu$_{90}$Co$_{10}$, however, prepared under
different conditions\cite{prb_56_1997_6086}. With
these values and using Eqs.\ref{lognormal} and
\ref{q_momento}, we could obtain the $q$
parameter for both samples: (a) $q= 3.11$ and (b)
$q=2.90$.

With those values of $q$, the non-extensive magnetic
susceptibility was obtained:
\begin{equation}\label{sus_generalizada_1}
\chi_q=\frac{q\mu_{ne}^2}{3kT}=\frac{q(2-q)^2\langle\mu\rangle^2}{3kT}
\end{equation}
and matches the experimental one, as presented in
figure \ref{reisfig2}-bottom.

\subsection{Distribution of critical temperatures}

Campillo and co-workers\cite{jmmm_237_2001_61} analyzed the
magnetic properties of 200-nm thick films of
La$_{0.67}$Ca$_{0.33}$MnO$_3$, which were grown under identical
conditions onto five different single crystal substrates: MgO, Si,
NdGaO$_3$, SrTiO$_3$, LaAlO$_3$. The films exhibit a strong
substrate dependence of the magnetic properties, including, for
instance, different values of Curie temperature $T_C$. On the
other hand, the strain due to the growth mode induces
inhomogeneities on the
sample\cite{jmmm_237_2001_61,nature_428_2004_401}. Campillo
verified that these inhomogeneities depend on the substrate
character and therefore they determined the $T_C$ distribution for
each thick film.

\begin{figure}
\begin{center}
\includegraphics[width=7cm]{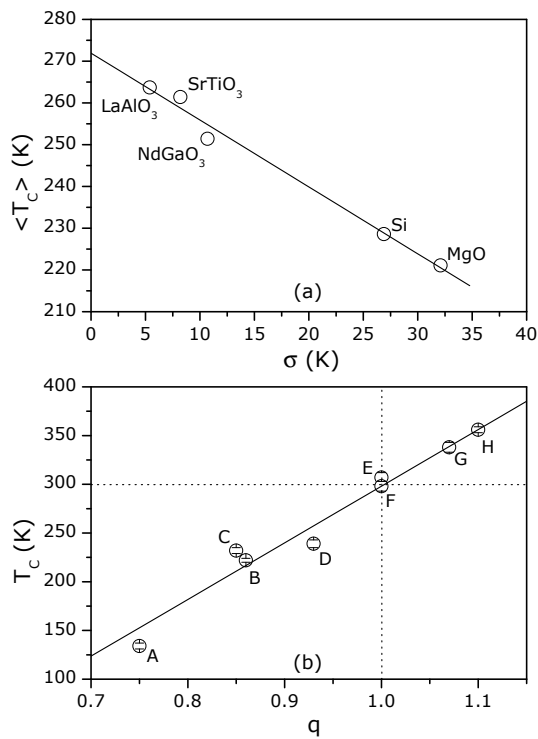}
\end{center}
\caption{(a) First moment $\langle T_C\rangle$ of
the Curie temperature distribution as a function
of the standard deviation $\sigma$, obtained
fitting the $M$ vs. $T$ curves of
La$_{0.67}$Ca$_{0.33}$MnO$_3$ manganites
deposited onto five different single crystal
substrates: MgO, Si, NdGaO$_3$, SrTiO$_3$,
LaAlO$_3$. After Campillo et al.
\cite{jmmm_237_2001_61}. (b) Curie temperature as
a function of the $q$ parameter, obtained from
the fit, using the generalized Brillouin
function\cite{PRB_66_2002_134417,EL_58_2002_42},
of the $M$ vs. $T$ curves of those manganites
referred in Table \ref{tab1}.} \label{reisfig3}
\end{figure}

The authors considered that the magnetization of a small and
homogeneous region is given by:
\begin{equation}\label{magcrit}
    \mathcal{M}(m_0,T_C,T,\beta)=m_0\left(1-\frac{T}{T_C}\right)^\beta\theta(T_C-T)
\end{equation}
where $\theta(x)$ is the Heavyside Step function. Supposing a
Normal distribution of Curie temperatures
\begin{equation}\label{ftc}
    f(T_C)=\frac{1}{\sqrt{2 \pi}\;\sigma}\;\exp\left(-\frac{\left(T_C-\langle T_C\rangle\right)^2}{2\;\sigma^2}\right)
\end{equation}
where $\sigma$ is the standard deviation and $\langle T_C \rangle$
the first moment of the distribution, they could write the average
magnetization:
\begin{equation}\label{mag_media_momento_2}
\langle\mathcal{M}\rangle=\int_0^\infty
\mathcal{M}(m_0,T_C,T,\beta)\;f(T_C)dT_C
\end{equation}
and fit the corresponding $M$ vs. $T$ curves, for
all samples available. From those fits, the
authors could obtain a linear relationship
between the standard deviation $\sigma$ and the
mean value $\langle T_C\rangle$ of the samples
analyzed; $\langle T_C\rangle=272-1.6 \;\sigma$,
as presented in figure \ref{reisfig3}(a).

On the other hand, in our previous
work\cite{EL_58_2002_42}, we used the Generalized
Brillouin
function\cite{PRB_66_2002_134417,EL_58_2002_42},
a non-extensive magnetic equation of state, to
fit the $M$ vs. $T$ curves obtained from those
manganites presented in Table \ref{tab1}. From
that study emerges a linear relationship between
$T_C$, a quantity intrinsic to the material, and
the $q$ parameter, a quantity intrinsic to the
non-extensive statistics; $T_C=-283+581\; q$, as
presented in figure \ref{reisfig3}(b).
\begin{table}[tbp]
\caption{Manganites, and the respective
references, in which its $M$ vs. $T$ curves were
analyzed using the generalized Brillouin
function\cite{PRB_66_2002_134417,EL_58_2002_42}
in order to obtain the corresponding $q$
parameter. The Label column corresponds to the
figure \ref{reisfig3}(b). The choice of compounds
was made such as to cover a wide range of $T_c$
values within the ferromagnetic phase, but it was
random in any other aspect.} \label{tab1}
\begin{tabular}{|c|c|c|}
\hline Label & Compound & Ref. \\ \hline\hline A &
La$_{0.62}$Y$_{0.07}$Ca$_{0.31}$MnO$_{3+\delta }$ &
\cite{JAP_83_1998_7076} \\ \hline B &
La$_{0.875}$Sr$_{0.125}$MnO$_{3+\delta }$ &
\cite{PRB_54_1996_6172} \\ \hline C &
La$_{0.5}$Ca$_{0.5}$MnO$_{3}$ & \cite{PRB_62_2000_6437} \\ \hline
D & La$_{0.83}$Sr$_{0.17}$Mn$_{0.98}$Fe$_{0.02}$O$_{3}$ &
\cite{PRB_57_1998_8509} \\ \hline E &
La$_{0.89}$Sr$_{0.11}$MnO$_{3+\delta }$ & \cite{EL_58_2002_42}\\
\hline F & La$_{0.75}$Ba$_{0.25}$MnO$_{3}$ &
\cite{JMMM_219_2000_1} \\\hline G & La$_{0.5}$Ba$_{0.5}$MnO$_{3}$
& \cite{JMMM_219_2000_1} \\ \hline H &
La$_{0.7}$Sr$_{0.3}$Mn$_{0.9}$Ru$_{0.1}$O$_{3}$ &
\cite{APL_77_2000_2382} \\ \hline
\end{tabular}
\end{table}

Comparing the results obtained by our
own\cite{EL_58_2002_42} (figure
\ref{reisfig3}(b)) and
Campillo\cite{jmmm_237_2001_61} (figure
\ref{reisfig3}(a)), we could estimate a relation
between the standard deviation $\sigma$ and the
$q$ parameter:
\begin{equation}\label{tcvsq}
\sigma\propto(1-q)
\end{equation}
reinforcing the idea that the $q$ parameter is related to the
inhomogeneities of the system. Note that, $q\rightarrow$1 imply in
$\sigma\rightarrow$0, i.e., the extensive (Maxwell-Boltzmann)
limit corresponds to the homogeneous case.

\section{Concluding remarks}

Summarizing, in the present work we shown that the $q$ parameter
measures the inhomogeneity and dynamics of a given inhomogeneous
magnetic system. We stress that the measured $q$, obtained from
scanning tunnelling spectroscopy on manganites (a microscopic
information), is able to predict a thermodynamic quantity, the
bulk magnetization (a macroscopic information); the entropic
parameter contains the dynamics, connecting the microscopic and
macroscopic worlds. Thus, the present work shows a clear
combination of dynamics and statistics, the key role to describe
complex systems. The present model was also successful applied to
a series of La$_{0.67}$Ca$_{0.33}$MnO$_3$ thick
films\cite{jmmm_237_2001_61} and melt-spun Cu$_{90}$Co$_{10}$
ribbons\cite{prb_56_1997_6086}; and the results reinforce the
conclusions made in this work.

\section{Acknowledgements}

We acknowledge CAPES-Brasil and GRICES-Portugal.
MSR acknowledges the FCT Grant No.
BPD/23184/2005. We are also thankful to P.B.
Tavares and A.M.L. Lopes (sample preparation),
and M.P. Albuquerque and A.R. Gesualdi (some
discussion on image processing).

\end{document}